# The 2009 Outburst of V630 Cassiopeiae

**Jeremy Shears and Gary Poyner**

**Abstract**

We present observations and analysis of the 2009 outburst of the unusual dwarf nova V630 Cas which is only the third recorded outburst of this star. The outburst lasted about 104 days, with the rise to maximum being slightly slower than the decline, which we interpret as an inside-out outburst. At is brightest it had V = 14.0, 2.3 magnitudes above the mean quiescence magnitude. The characteristics of the outburst are similar to several other long orbital period dwarf novae.

**Introduction**

V630 Cas is a cataclysmic variable (CV) star that was first observed to undergo an outburst in 1950 lasting about 2 months during with the star reached $m_{pg}$ = 12.3 [1]. A second outburst was observed in 1992 with a 70 day rise to maximum at V = 14.3 and a decline over about 30 days [2]. Based on its similarity to BV Cen, GK Per and V1017 Sgr, Warner [3] suggested that V630 Cas is an unusual dwarf nova of long orbital period. Orosz *et al*. [4] presented extensive spectroscopy and photometry of the star, confirming it is indeed a dwarf nova with a very long orbital period of 2.56387(4) days and an inclination in the range 67 to 78º. Furthermore, the long term light curve during quiescence showed variations of up to 0.4 mag on short (1 – 5 d) timescales and variations of 0.2 – 0.4 mag on longer timescales (3 – 9 month). They concluded that the secondary star is a "stripped giant" of late spectral type K4-5 and a mass $M_2$ ~ 0.17 $M_{sun}$ with the white dwarf $M_1$ ~ 0.98 $M_{sun}$.

In this paper we present the light curve of the 2009 outburst of V630 Cas, only the third recorded outburst, and compare it with the 1992 and 1950 outbursts.

**Observations**

An image of V630 Cas in outburst is shown in Figure 1. The light curve of the 2009 outburst is shown in Figure 2 using data submitted to the BAA VSS Recurrent Objects Programme, covering the year from 2008 June. In mid February 2009, several observers noticed that V630 Cas was showing a brightening trend. As a result, a specific request for observations was made via the baavss-alert and cvnet-outburst email lists on 2009 Feb 18 [5, 6]. Compared with other campaigns on CVs, relatively few observers participated, most likely as the field is not well positioned in the evening sky during this period. Observations were contributed by the following:

Visual:  G. Poyner
CCD unfiltered:  R. Januszewski, H. McGee (observations made with a "GRAS Global Rent-a-Scope" telescope), I. Miller, M. Mobberley, R. Pickard, J. Shears (including observations made with the Bradford Robotic Telescope)

CCD+V: D. Boyd, I. Miller, M. Mobberley, M. Nicholson, G. Poyner (including observations made with the Sonoita Research Observatory telescope)

CCD+R: D. Boyd, G. Poyner (including observations made with the Sonoita Research Observatory telescope)

**The course of the 2009 outburst**

Determining the precise start and end of such a slow outburst is difficult, especially since during quiescence there is an underlying variation of around 0.4 magnitudes. We established the average magnitude at quiescence as 16.3 based on the observations between JD 2454470 and 2454850, i.e. prior to the outburst. Thus if we assume that outburst is when the star is above 16.0, the outburst started on or around 2009 Jan 24 (JD 2454856) and ended around 2009 May 9 (JD 2454960), a total of 104 days. At its brightest it reached V = 14.0 and the time of maximum, Mar 26 (JD 2454917), was determined by fitting a cubic function to the upper part of the light curve. Thus the outburst amplitude was 2.3 magnitudes above quiescence. The outburst profile is approximately symmetrical, although the rise to maximum, 61 days, was slower than the decline, at 43 days. Such a profile is indicative of an *inside-out* outburst in which the thermal instability which triggers the outburst starts at the inner edge of the accretion disc and propagates towards the outer edge. As we shall see below, this is typical of long orbital period dwarf novae, but is in contrast to most dwarf nova outbursts in which the rise is much faster than the decline in which the outburst starts at the outer edge of the disc [7].

**Comparison with earlier outbursts**

Light curves of the 2009, 1992 and 1950 outbursts are shown in Figure 3, all drawn to the same scale for ease of comparison. Our measurements of the maximum magnitude ($m_{max}$), amplitude, rate of rise ($\tau_r$), rate of decline ($\tau_d$), the number of days to maximum and the outburst duration are shown in Table 1. Unfortunately the decline phases of the two earlier outbursts were poorly observed, which means the decline rate and outburst duration are not well constrained.

The 2009 and 1992 outbursts appear to be rather similar in terms of their maximum brightness, rate of rise, the time taken to reach maximum and the outburst duration. However the 2009 outburst shows a single well-defined maximum, but this is not the case for the 1992 outburst which appears broader and possibly double peaked [2]. By contrast, the 1950 outburst appears to be rather different from the other two, being significantly brighter and with a much faster rise rate. Although rather few data points were obtained during the decline from the 1950 outburst, the limits on the decline rate suggest it, too, was faster. Thus all three outbursts appear to have faster declines than rise, consistent with them all being inside-out outbursts.

Bailey [9] found that $\tau_d$ of dwarf novae is well correlated with $P_{orb}$ and Warner [10] presented a relationship between the two parameters (reference 10, equation 3.5). We have used the same data as Warner and supplemented it with our measured $\tau_d$ values from the 2009 outburst of V630 Cas and another long period dwarf nova, SDSS J204448.92-045928.8 [11, 12] to refine the relationship between $\tau_d$ and $P_{orb}$ for dwarf novae. The correlation is shown in Figure 4 and an exponential fit to the data gives the following relationship:

$$\tau_d = 0.579 \, P_{orb}^{0.782} \text{ (h) d/mag} \qquad \text{(Equation 1)}$$

**Comparison with other long $P_{orb}$ CVs**

As noted in the Introduction, V630 Cas is an example of a CV with a very long orbital period. Only 6 CVs, including V630 Cas, are currently known to have $P_{orb}$ above 1.5 days [13] and these are listed in Table 2. Of these, four are known to undergo dwarf nova-type outbursts: V1017 Sgr, V630 Cas, GK Per and SDSS J204448.92-045928.8. These are slow affairs, commensurate with their long $P_{orb}$ and large accretion discs, and of modest amplitude: 2 to 4 mags. The system with the longest known $P_{orb}$ is V1017 Sgr at 5.7 d. This is a recurrent nova, initially known as N Sgr 1919, that has also exhibited several dwarf nova outbursts approximately every 18 years [14]. Recently, SDSS J204448.92-045928.8 was reported for the first time to have undergone a dwarf nova-type outburst lasting about a month [12].

GK Per was first observed as N Per 1901 and it is the best characterised in terms of the number of dwarf nova outbursts of all the long $P_{orb}$ CVs. These outbursts only became prominent after 1947, more than 45 years after the nova explosion. Simon [15] presented an analysis of 20 outbursts and found that although the amplitude is typically ~3 magnitudes, it has varied by more than 1.5 magnitudes. However, after 1982 and up to the end of the period analysed, 2000, the amplitude has stabilised. In the case of V630 Cas although we only have the three known outbursts to consider, but we note that its amplitude has also varied by 2 magnitudes. The variation in the maximum brightness of both systems, which represents a considerable fraction of the overall amplitude, may depend on how much of the accretion disc goes into the hotter state [16]. In such systems with very large accretion discs, the slow moving heating front may not always propagate from the inner edge of the disc all the way to the outer edge before it meets the cooling wave which starts at the outer edge and moves inward.

The majority of GK Per's outburst light curves are almost symmetrical and are probably of the inside-out variety [16], as with V630 Cas. However, several anomalous outbursts of GK Per have been observed. For example, the 2006 outburst was 1.5 mags fainter than normal and its light curved appeared to be triple-peaked, suggestive of 3 short, faint outbursts running into each other [17]. This led Evans *et al.* [17] to propose that the heating wave, triggered at the inner accretion disc passes through the disc, but at some point encounters a lower-density region, halting its progress. The cooling wave then moves inwards reducing the amount of the disc which is in the hot state. However, when it reaches the lower density region, it falters and thus fails to fully extinguish the outburst. The inner disc still remains hot and rekindles the outburst and the cooling wave is reflected back as a heating wave, resulting in a second (and ultimately a third) rebrightening [17]. If this idea is correct, it could also explain the apparent double-peaked nature of the 1992 outburst of V630 Cas.

GK Per and V1017 Sgr are rare examples of CVs which have been observed to exhibit *both* nova and dwarf nova outbursts. Menzies et al. [18] also suggested that BV Cen, another dwarf nova with a long orbital period, $P_{orb}$ = 0.61 d [13], may have undergone an unrecorded nova outburst based on its similarities with GK Per. Furthermore, Sekiguchi [14] noted that BV Cen, GK Per and V1017 Sgr not only have similarities in their dwarf nova outbursts light curves, but also show very similar outburst spectra, having a blue continuum and strong HeII, 4640Å, HeI and Balmer emission lines. He went on to suggest that these three CVs constitute a class of long $P_{orb}$ "hybrid classical/dwarf novae" which may represent a sequence of evolution of the same type of object. It is therefore tempting to speculate that V630 Cas could be a member of this class. We note that all systems have similar almost symmetrical dwarf nova outburst profiles, amplitudes, long recurrence times, long rise times to maximum (roughly correlated with the system's $P_{orb}$). Unfortunately, we are not aware of spectroscopy being conducted on V630 Cas during any of its outbursts, which could provide further support for these CVs being related. We also note that BV Cen, GK Per and V1017 Sgr are hard X-ray sources [14], but neither are we aware of any X-ray measurements on V630 Cas. Clearly spectroscopy and X-ray observations during a future outburst are a priority which could help confirm whether or not it is related to the other three systems.

Obviously, if V630 Cas were a "hybrid classical/dwarf novae", it should have undergone a nova explosion sometime in the past, possibly reaching magnitude 3 to 7, but there is no evidence that this is the case, at least in historical times. We have examined deep images of the field of V630 Cas, e.g. from the Palomar Observatory Sky Survey, but find no evidence of a nova shell such as exists around GK Per. Of course, such a shell could have dispersed if the nova explosion occurred a sufficiently long time in the past. Thus whether V630 Cas is such a hybrid system can only remain speculation at this time.

**How often does V630 Cas outburst?**

Most dwarf novae outburst quasi-periodically. For example V1017 Sgr outbursts approximately every 18 years [14]. By contrast, the outburst period of GK Per has evolved from 385 days in the years 1948 to 1967, to the more recent situation where the period has been much longer, ~1000 days, and is gradually increasing. We consider it unlikely, although not impossible, that an outburst of V630 Cas has been missed between the ones in 1992 and 2009. As we have seen, the outbursts are long lasting and it has received intense coverage by the Indiana Automated CCD Photometric Telescope covering the thirteen years from the previous outburst until 2005 February [8] and in subsequent years there has been intense monitoring by members of the BAA and AAVSO. If this 17 year separation represents the typical outburst period, it is possible that an outburst was missed between those in 1950 and 1992, perhaps between 1967 and 1975 (i.e. 17 years from the known outbursts) when the star received much less, if any, observational attention – for example, no observations exist in the AAVSO international database during this interval. Based on the 17 year gap between the last two outbursts, we may have to wait until 2026 until the next outburst. However, predicting dwarf nova outbursts is notoriously difficult, hence we encourage all observers to keep this intriguing star under regular observation.

**Conclusions**

We have presented and analysed light curve of the outburst of the unusual dwarf nova V630 Cas which was observed in early 2009. This was only the third recorded outburst of this star, the previous ones being in 1950 and 1992. The outburst lasted about 104 days, with the rise to maximum, at 61 days, being slightly slower than the decline, 43 days, which we interpret as an inside-out outburst. At is brightest it had V = 14.0, 2.3 magnitudes above the mean quiescence magnitude. The outburst profile was similar to the 1992 outburst, but was significantly fainter than the one in 1950. We used our measurements of the decline rates, $\tau_d$, of the 2009 outburst and of another long $P_{orb}$ dwarf nova, SDSS J204448.92-045928.8, to refine the previously known relationship between $\tau_d$ and $P_{orb}$ for dwarf novae.

The characteristics of the 2009 outburst are similar to those of several other long orbital period dwarf novae, notably V1017 Sgr, GK Per and BV Cen. We suggest that V630 Cas could be a hitherto unrecognised member of the "hybrid classical/dwarf novae" group, although there is no evidence for a nova outburst having occurred.

**Acknowledgements**

The authors are most grateful to the following for contributing their observations of V630 Cas to the campaign: D. Boyd, R. Januszewski, H. McGee, I. Miller, M. Mobberley (who also gave permission to use his image as Figure 1), M. Nicholson and R. Pickard. We are deeply indebted to Professor R. Kent Honeycutt (Astronomy Department, Indiana University, Bloomington, USA) for allowing us to use his photometry of the 1992 outburst, as well as for his encouragement. JS acknowledges access to the Bradford Robotic Telescope operated by the Department of Cybernetics, University of Bradford, UK, which was used to obtain unfiltered CCD images. GP acknowledges the AAVSO and the Sonoita Research Observatory for V and R band CCD observations.

**Addresses:**
JS: "Pemberton", School Lane, Bunbury, Tarporley, Cheshire, CW6 9NR, UK
[bunburyobservatory@hotmail.com]

GP: 67 Ellerton Road, Kingstanding, Birmingham, B44 0QE, UK
[Garypoyner@blueyonder.co.uk]

| Outburst year | $m_{max}$ | Amplitude [a] mag | $T_r$ [b] d/mag | $T_d$ [b] d/mag | Time to maximum d | Outburst duration [c] d |
|---|---|---|---|---|---|---|
| 2009 | V = 14.0 | 2.3 | 28 | 19 | 61 | 104 |
| 1992 | V = 14.3 | 2.0 | 33 | <23 | 63 | 120 |
| 1950 | $m_{pg}$ = 12.3 | 4.0 | 17 | <7 | >17 | >74 |

**Table 1: Outburst parameters for three outbursts of V630 Cas**

[a] Assuming a quiescence magnitude of 16.3

[b] Measured from the gradient of the light curve midway between quiescence at maximum

[c] Assuming any measurement of the star at mag 16 or brighter is in outburst

| Object | Type | $P_{orb}$ d | Spectral type of secondary | Typical Dwarf nova outburst amplitude mag | Rise time d | Recurrence time |
|---|---|---|---|---|---|---|
| V1017 Sgr | Recurrent nova + Dwarf nova | 5.714 | G5/3p | 3 [14] | ~100 [14] | ~18 yr [14] |
| MR Vel | Super soft X-ray source | 4.028782 | | | | |
| V630 Cas | Dwarf nova | 2.56387 | K4-5 | 2.3-4 | ~60 | 17-42 yr |
| GK Per | Nova + Dwarf nova | 1.996803 | K1/4 | 3 [15] | ~30 [15] | ~1000 d [15] |
| J0305+0547 N Cet 2008 | Nova | 1.77 | | | | |
| SDSS J204448.92-045928.8 | Dwarf nova | 1.68 | K4-5 | 2 | ~15 | |

**Table 2: Properties of long $P_{orb}$ CVs**

Confirmed dwarf novae are highlighted in grey. $P_{orb}$, spectral type of the secondary are from reference 13. Additional data for V630 Cas are from the present study and for SDSS J204448.92-045928.8 are from our measurements on the outburst profile shown in reference 12

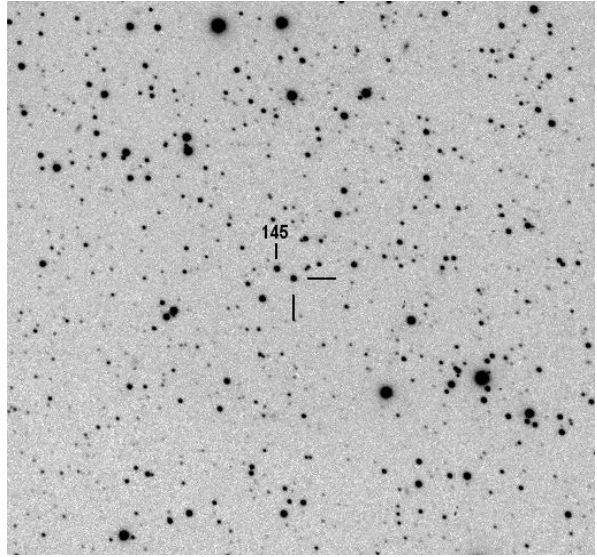

**Figure 1: Image of V630 Cas in outburst on 2009 Mar 18.826 at mag 14.6**

*(Martin Mobberley)*

N at top, E to left, 13 min x 13 min. Comparison star 145 has V = 14.485. Equipment: 0.35 cm f/7 SCT, SBIG ST9XE CCD unfiltered camera, 120 s exposure

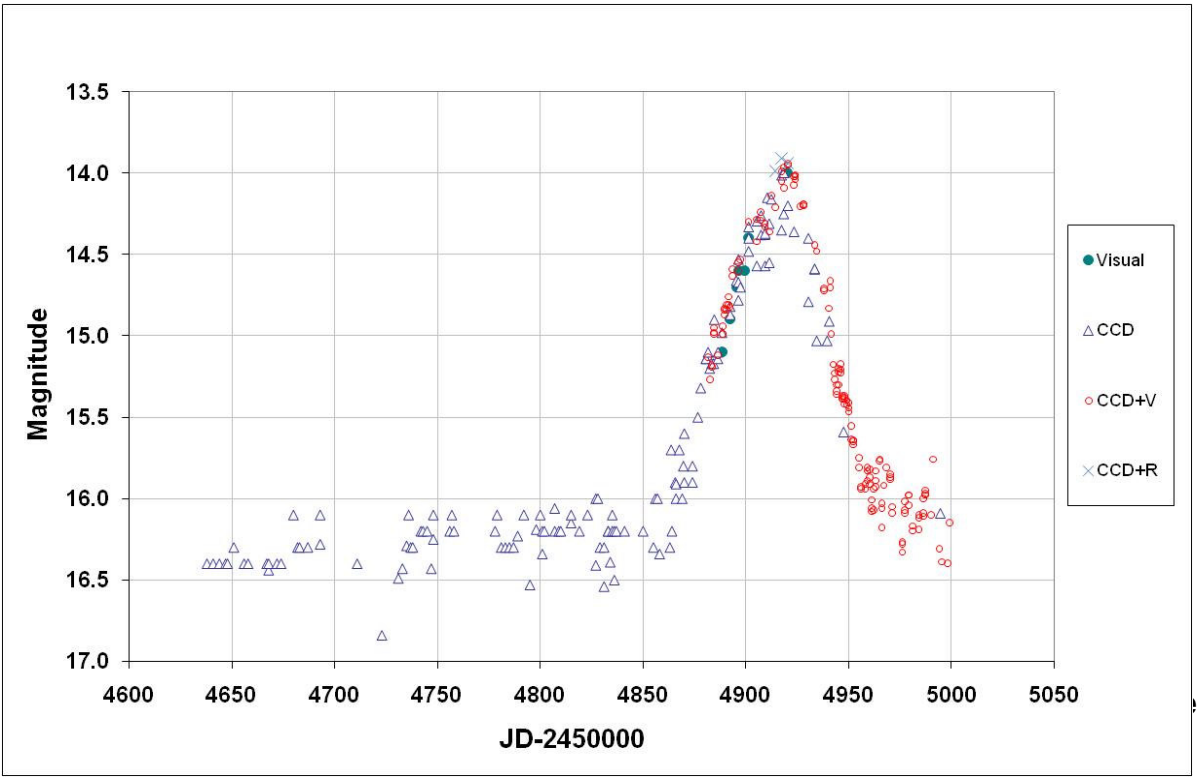

Figure 2: Light curve of V630 Cas between 2008 June 20 to 2009 June 16

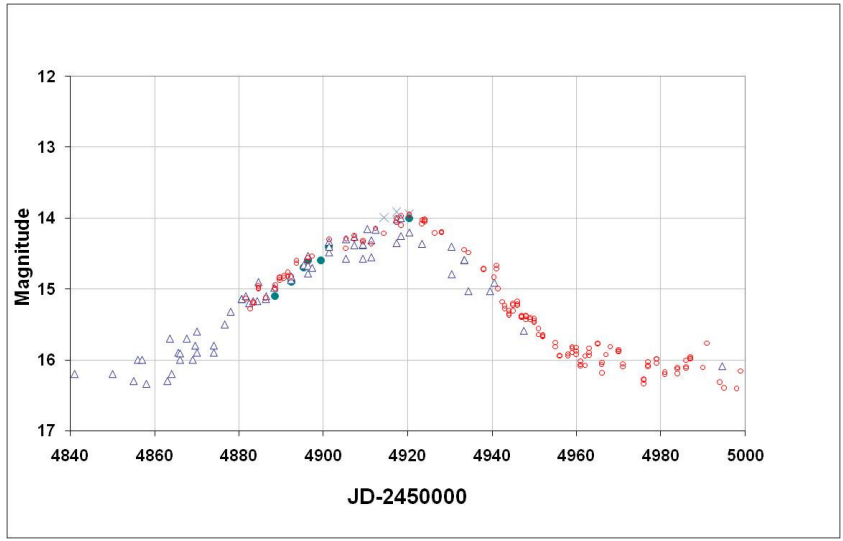

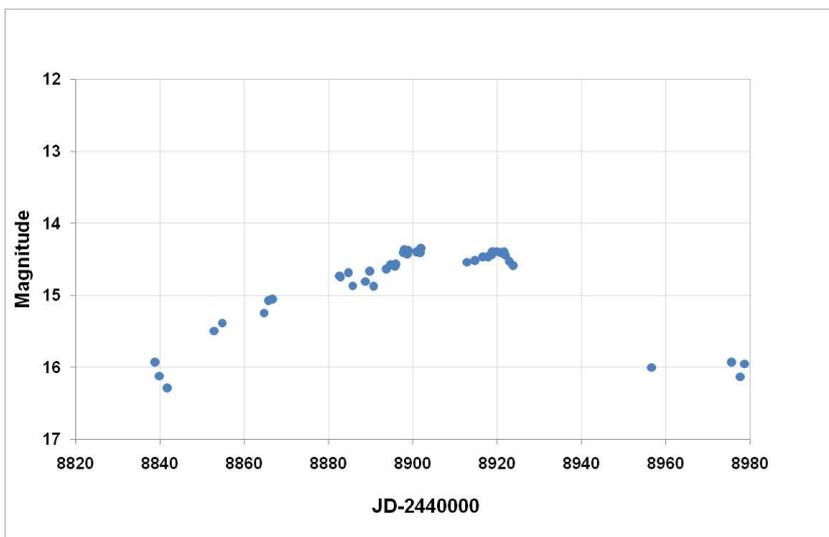

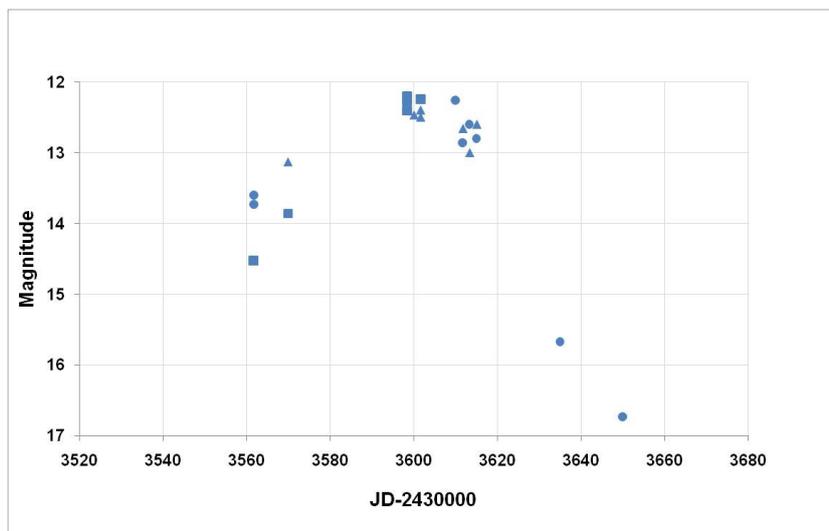

**Figure 3: Light curves of the 2009 (top), 1992 (middle) and 1950 (bottom) outbursts**

2009 data are from Figure 1. 1992 data are V-band photometric measurements from the Indiana Automated CCD Photometric Telescope, supplied by Kent Honeycutt [8]. 1950 data are measured from the Figure in Reference 1: ■ green sensitive plate, camera lens; ● green sensitive plate, reflector; ▲ red sensitive plate, reflector

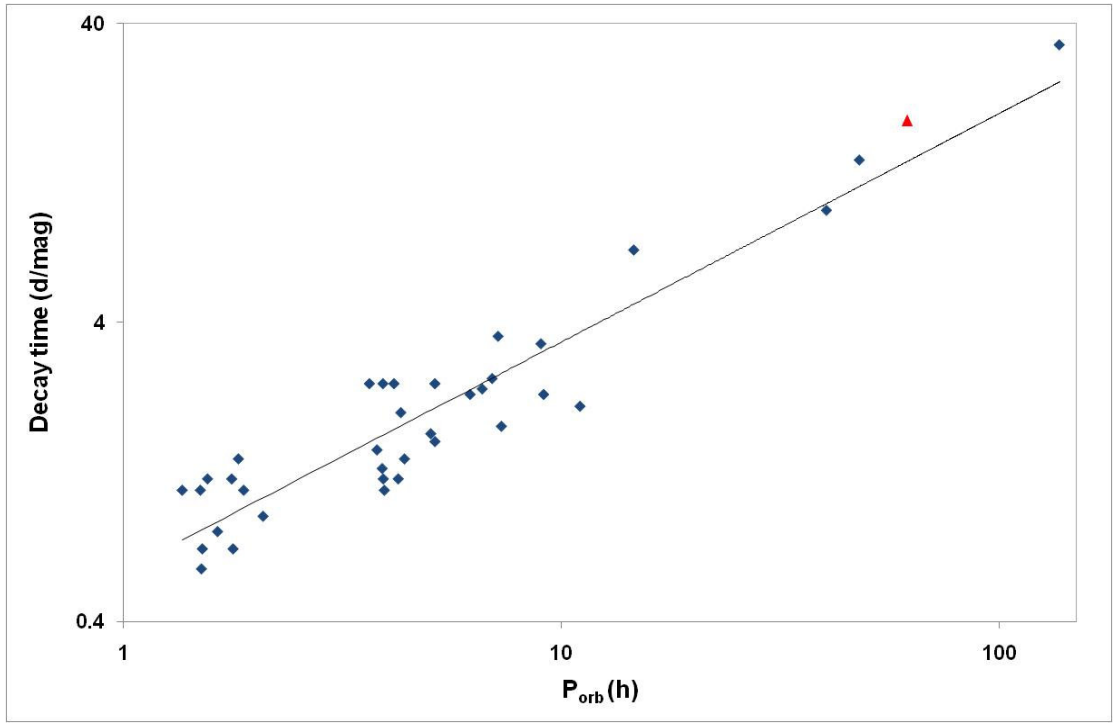

**Figure 4: Decay time scale, $\tau_d$, of dwarf nova outbursts**

The red triangle denotes data for the 2009 outburst of V630 Cas from the current study. Other data are from reference 10, Tables 3.1 to 3.3, supplemented with our own measurement of $\tau_d$ for SDSS J204448.92-045928.8 from the light curve presented by in reference 12